\documentclass[preprint,nofootinbib,eqsecnum,floats,aps]{revtex4}
\usepackage{graphicx}
\usepackage{bm}
\usepackage{epsfig}

\def\bnu{\begin{enumerate}}
\def\enu{\end{enumerate}}
\def\be{\begin{equation}}
\def\ee{\end{equation}}
\def\br{\begin{eqnarray}}
\def\er{\end{eqnarray}}
\def\brn{\begin{eqnarray*}}
\def\ern{\end{eqnarray*}}
\def\go{\rightarrow}

\def\ie{{\em i.e., }}
\def\nn{\nonumber }
\def\rf#1{{(\ref{#1})}}

\def\b{\beta}

\def\l {{\lambda}}

\def\ket#1{|#1 \rangle}

\def\etal{{\it et al }}

\begin{document}

\title{Pairing Correlations in Odd-Mass Carbon Isotopes and Effect of
 Pauli  Principle in Particle-Core Coupling in $^{13}$C and  $^{11}$Be}
\author{A.R. Samana$^{1,2}$}
\author{T. Tarutina$^3$}
\author{F. Krmpoti\'c$^{3,4,5}$}
\author{M.S. Hussein$^2$}
\author{T.T.S. Kuo$^6$}

\affiliation{$^1$Centro Brasileiro de Pesquisas F\'{\i}sicas,
CEP 22290-180, Rio de Janeiro, RJ, Brazil}

\affiliation{$^2$Facultad de Ciencias Exactas
, Departamento de F\'isica, Universidad Nacional de La Plata, 1900 La Plata,
Argentina}

\affiliation{$^3$Departamento de F\'isica Matem\'atica, Instituto
de F\'isica da Universidade de S\~ao Paulo,
Caixa Postal 66318, 05315-970 S\~ao Paulo, SP, Brazil}

\affiliation{$^4$Instituto de F\'isica La Plata, CONICET, 1900 La
Plata, Argentina}

\affiliation{$^5$Facultad de Ciencias Astron\'omicas y
Geof\'isicas, Universidad Nacional de La Plata, 1900 La Plata,
Argentina}

\affiliation{$^6$Department of Physics and Astronomy,
State University of New York, Stony Brook, New York, 11794-3800, USA}

\date{\today}
\begin{abstract}
We present  an exploratory study of
structure of  $^{13}$C,  $^{15}$C, $^{17}$C and
 $^{19}$C, showing that  the simple  one-quasiparticle  projected  BCS
 (PBCS) model is capable to account
for several important~ pro-\\perties  of these nuclei.
Next we discuss the importance of the Pauli Principle
in the particle-core models of normal-parity states in $^{13}$C and
$^{11}$Be. This is done by considering the pairing interaction
between nucleons moving in an over-all deformed potential.
To assess the importance of  pairing correlations in
these light nuclei we use  both the simple BCS and the
PBCS approximations.
We show that the Pauli Principle plays a crucial role in the parity
inversion in $^{11}$Be.
It is also found that the effect of
the particle number conservation in  relatively light and/or exotic
nuclei is quite significant.
 Comparison of our results with several recent papers on the same
subject, as well as with some experimental data,  is  presented.
\end{abstract}


\maketitle

\section{Introduction}

It has been known for a long time that the structure of the
nucleus depends significantly on their superfluid nature. In fact,
pairing constitutes the main part of the residual interaction
beyond the Hartree-Fock (HF) approximation and has a strong
influence on most low-energy properties of the
system~\cite{bender03b}. This encompasses masses, separation
energies, deformation, individual excitation spectra and
collective excitation modes such as rotations and  vibrations. The
role of pairing correlations is particularly emphasized when going
toward the neutron drip-line because of the proximity of the Fermi
surface to the single-particle continuum. Indeed, the scattering
of virtual pairs into the continuum gives rise to a variety of new
phenomena in ground and excited states of nuclei~\cite{doba3}.

The above suggests that the  pairing correlations might play a relevant role
in the structure of light exotic nuclei.
However, the fact that   the  $^{11}$Be ground state is a
 ${1}/{2}^+$ state  and not a ${1}/{2}^-$ state  (as dictates the
spherical shell  model)
 is frequently attributed to the quadrupole core excitation
effects only.  That is, the parity inversion in $^{11}$Be
 is usually described within the simple
particle-vibration coupling model (PVM) \cite{Vin95,Col01,Gor04,Bro04},
or within the simple particle-rotor coupling model (PRM)
 \cite{Esb95,Nun95,Nun96,Rid98},
where  the correlations  among the valence
particles, coming from the residual interaction,  are totally neglected
\footnote{In the early 1960's Talmi and Unna~\cite{Tal60}   first
  noticed the parity inversion  in $^{11}$Be and suggested that the
  interaction of a $1p_{1/2}$ or   $2s_{1/2}$ neutron with two   $1p_{3/2}$
 protons was the main cause. Many large-basis shell-models
 calculations have been carried out for $^{11}$Be afterwards, but they do not
all reproduce the parity inversion. Moreover,
a recently done large basis ab initio shell model
study~\cite{For05}  suggests that  a realistic 3N force
will have an important influence on this regard.}.

We know, however, that, while the positive parity states in $^{11}$Be
 can be accounted for fairly well in the weak
coupling model \cite{Esb95}, the low-lying negative parity states
cannot be described reliably with the same model. As
stated by  Esbensen \etal \cite{Sag93},  this is because the last have
a complicated structure
due to the importance of the pairing correlations.
The same statement is also valid for the  $^{13}$C nucleus.
In fact, it has been recognized since the paper of  Lane~\cite{Lan55}  on $^{13}$C
 in the mid-50's that the weak-coupling of a $sd$  nucleon to a $p$-shell core is a good
starting point for the description of low-lying non-normal-parity states
in a number of $p$-shell nuclei. Moreover, Lawson and Kurath
 have  also pointed out long ago~\cite{Kur61} that the same is not true for the
normal-parity states:
''The reason is that the single-nucleon function which is added to the $^{12}$C core
has strong components in common with some of the functions within the core.''
Thus, because of the violation of the Pauli Principle (PP) the particle-core weak-coupling
models have been mostly limited to the study the positive parity spectrum
of $^{13}$C~\cite{Kur61,Mik71,Rob72,Esb95}.
Although important  efforts have been invested to incorporate the PP into the
core-particle models~\cite{Gor04,Bro04,Bar01},  the shell-model was so far the
only  plausible alternative to  deal with the negative parity
states in odd-mass carbon and $^{11}$Be nuclei~\cite{Coh67,Suz03}.

But the pairing correlations are closely related to
the PP, which is not violated
within the BCS approximation, the fact that is not always well perceived.
The problem that appears now, however, is
the well  known particle-number non-conservation. This, in turn,
can be circumvented through the particle-number projection procedure
~\cite{Ott69,Krm93}, \ie in the framework of the projected BCS (PBCS)
approximation.
In fact, recently it has been found~\cite{Krm02,Krm05} that the low-lying
energy spectra in  $^{13}$C can be described  quite well
in the context of  the PBCS approximation
for  the pairing force among the valence neutrons.
Moreover,  in this study it was shown that the projection
procedure is very important to  account for  the weak decay
observables around  $^{12}$C.

The aim of the present work is twofold. First, we
inquire to which extent  the main feature of  heavy carbon isotopes
$^{15}$C, $^{17}$C, and $^{19}$C
can be interpreted  within the framework of the PBCS model, using
as a building block the results obtained in Ref. \cite{Krm02,Krm05} for
 the stable carbon isotope $^{13}$C. In doing so, one should keep in
 mind that, as the number of neutrons increases, in going from
$^{13}$C to $^{19}$C,~ the nuclei become more and more weakly bound.
There is no systematic study of the energy spectra of
these nuclei in the literature.
Secondly,  we discuss   the interplay between the single-particle and
collective  degrees of freedom in  $^{13}$C and  $^{11}$Be, by
incorporating the PP into the
previous works \cite{Vin95,Col01,Nun95,Nun96,Esb95,Rid98}
through the  pairing correlations
\cite{Kiss63,Sch68,Ste71,Bar75}. Therefore, we will discuss
the structure of these two nuclei in the
quasiparticle-rotor model (QPRM) and the quasiparticle-vibrator
model (QPVM). The corresponding models with the particle number
projection included will be labelled, respectively, as PQPRM and PQPVM.
Special emphasis is put on the odd-parity states which so far have been
treated very rudimentary within the particle-core models.

The theoretical tools are certainly oversimplified as compared to the
physical reality of  nuclei that are discussed. In particular, neither the
weak-binding characteristics nor coupling to the continuum states is
taken into account. However, in spite of the simplicity of the approach
it allows for obtaining simple estimates of the main
structural features of these nuclei. Certainly, this
kind of calculation cannot replace any  full-fledged
description, which is, however, presently not available. The
present analysis has to be understood as a very simple version of
a shell-model approach, which nevertheless is able to give main
components of the involved configurations.
We also feel that the paper  makes some significant advance in the
line of research that has been quite importantly
 developed by previous similar studies.

\section{Pairing description of odd-mass carbon isotopes}

The definitions of particle  and hole states $E_{j}^{(\pm)}$ in
the  BCS and PBCS approximations are listed in Table~\ref{table1},
where
\be
{E}_{j}=({\bar e}_j^2+\Delta_j^2)^{1/2};\hspace{1cm}{\bar e}_j= e_j-\mu_j-\l,
\label{2.1}\ee
are the usual BCS quasiparticle energies,  which depend on the
 single-particle energies (s.p.e.) $e_j$, on the self-energy
\br
{\mu}_j&=& \sum_{j'}
{(2j'+1)^{1/2}\over (2j+1)^{1/2}} v_{j'}^2 {\rm
F}(jjj'j';0),
\label{2.2}\er
and on the pairing gap
\begin{eqnarray}
&&\Delta_j=-\frac{1}{2}\sum_{j'}{(2j'+1)^{1/2}\over
(2j+1)^{1/2}} u_{j'}v_{j'} {\rm G}(jjj'j';0).
\label{2.3}
\end{eqnarray}
\begin{table}[h]
\caption {Definitions of quasiparticle ($E_{j}^{(+)}$),  and
quasihole ($E_{j}^{(-)}$) energies in the BCS and PBCS
approximations. In both cases, $E_{j}^{(+)}$ can be either
 negative or positive, while $E_{j}^{(-)}$ are always negative.
 The BCS quasiparticle energies  $E_{j}$, defined in \rf{2.1}, are positive,
and  the corresponding chemical potential $\lambda$ is negative.
The  PBCS quasiparticle energies  $\varepsilon^{N}_j$ are defined in
\rf{2.4}. }
\begin{center}
\newcommand{\cc}[1]{\multicolumn{1}{c}{#1}}
\renewcommand{\tabcolsep}{2.7pc} 
\renewcommand{\arraystretch}{1.2} 
\label{table1}
\bigskip
\begin{tabular}{|c|c|c|}
\hline
 Model  & $E_{j}^{(+)}$& $E_{j}^{(-)}$\\
\hline\hline
BCS  &$\lambda+E_{j}$ &$\lambda-E_{j}$ \\ \hline
PBCS &$\varepsilon^{N}_j$ & $-\varepsilon^{N-2}_j$
 \\ \hline
\end{tabular}
\end{center}
\end {table}

The PBCS energies read
\br \varepsilon^{N}_j=
{R_0^N(j)+R_{11}^N(jj)\over I^N(j)}-\frac{R_0^N}{I^N}.
\label{2.4}\er
The quantities $R^N$  and $I^N$, where $N$ is the
neutron number, are defined in Ref. \cite{Ott69}.

The  BCS and PBCS predictions for the spectroscopic factors will be
also discussed.
Within the PBCS approximation these quantities
are given by \cite{Coh61,Ott69}:
\br
S_u(j) &=&u_j^2\frac{I^N(j)}{I^N},
\label{2.5}\er
for the stripping reactions on  even
targets and for pick up reactions on odd targets, and by
\br
S_v(j)&=&(2j+1)v_j^{2}\frac{I^{N-2}(j)}{I^N},
\label{2.6}\er
for the stripping reactions on  odd
targets and for pick up reactions on even targets.
The plain BCS results are recovered by making all the $I$-factors
equal to unity.

\begin{table}[h]
\caption{The quasihole energies $E_{j}^{(-)}$
with $j=1s_{1/2},1p_{3/2}$ and the quasiparticle energies $E_{j}^{(+)}$
with $j=1p_{1/2},\cdots,1f_{5/2}$, used
  in the fitting procedure,
and the resulting single-particle energies $e_j$, and the pairing
strength $v_{s}^{pair}$,   within the BCS and PBCS
for $N=6$. The energies
are given in units of MeV, and $v_{{s}}^{{pair}}$ is
dimensionless.}
\begin{center}
\label{table2}
\newcommand{\cc}[1]{\multicolumn{1}{c}{#1}}
\renewcommand{\tabcolsep}{2.4 pc}
\bigskip
\begin{tabular}{c| r| r r}\hline
$Shell$&$E_{j}^{(\pm)}$&$e_j(BCS)$&$e_j(PBCS)$
\\ \hline
$1s_{1/2}$&  $-35.00$       &$-23.29$& $-22.27$
\\
$1p_{3/2}$& $-18.72$&$-7.74$ &$-7.17$
\\
$1p_{1/2}$& $-4.94$ &$-2.01$ &$-1.44$
\\
$1d_{5/2}$& $-1.09$ &$2.18$  & $2.22$
\\
$2s_{1/2}$& $-1.85$ &$2.78$  & $2.74$
\\
$1d_{3/2}$& $3.26$  &$6.88$  & $6.89$
\\
$1f_{7/2}$& $8.63$  &$11.04$  & $11.06$
\\
$2p_{3/2}$& $7.24$  &$11.61$ & $11.62$
\\
$2p_{1/2}$&  $12.97$       &$17.43$ & $17.45$
\\
$1f_{5/2}$& $16.78$        &$19.27$ & $19.31$
\\ \hline
$v_{{s} }^{{pair}}$&  &$23.46$ &$24.22$\\\hline
\end{tabular}
\end{center}
\end{table}

The BCS and PBCS calculations presented  here were
performed in the same way as in the
previous works \cite{Krm02,Krm05}. That is,
for the residual interaction we adopted  the delta force,
\br
V &=&-4 \pi v^{pair}\delta(r)~\mbox{MeV-fm}^3,
 \label{2.7} \er
and the  configuration space includes the neutron  orbital with
$j\equiv nlj=(1s_{1/2},1p_{3/2},1p_{1/2},1d_{5/2},2s_{1/2},1d_{3/2}, 1f_{7/2},
1f_{5/2})$. The radial  wave functions were approximated by that
of the harmonic oscillator (HO) with the length parameter $b=1.67$
fm, which corresponds to the estimate $ \hbar \omega
=(45A^{-1/3}-25A^{-2/3})$~MeV for the oscillator energy.

The s.p.e. $e_j$, as well as the value of  the singlet pairing
strength $ v^{pair}$, were fixed by adjusting  the experimental
energies, taken from \cite[Table 13.4]{Ajz91}, to the calculated
ones, through a $\chi^2$ search assuming that: a) the ground state
${3/2}^{-}$ in $^{11}$C is the  ${1p_{3/2}}$ quasihole-state with
energy $E_{1p_{3/2}}^{(-)}$, and b) the lowest ${1/2}^{-},
{5/2}^{+},{1/2}^{+},{3/2}^{+}, {7/2}^{-}$, and ${3/2}^{-}$ states
in $^{13}$C are the  quasiparticle-states
$j=1p_{1/2},1d_{5/2},2s_{1/2},1d_{3/2}, 1f_{7/2}$, $2p_{3/2})$  with energies $E_{j}^{(+)}$.
The quasiparticle energies $E_{j}^{(\pm)}$ of the distant orbitals $1s_{1/2}$, $2p_{1/2}$
and $1f_{5/2}$, not known
experimentally, were assumed  to be: i) $E_{1s_{1/2}}^{(-)}$ MeV proposed in the study done by
Gillet and Vinh Mau \cite{Gil64}, ii) $E_{2p_{1/2}}^{(+)}=E_{2p_{3/2}}^{(+)}+\Delta_{ls}(p)$ MeV,
with $\Delta_{ls}(p)=5.73$ MeV being the spin-orbit splitting between
 the $p$-subshells, and iii)
 $E_{1f_{5/2}}^{(+)}=E_{2p_{3/2}}^{(+)}+\Delta_{HO}=16.78$
 MeV, were $\Delta_{HO}=3.82$ MeV is  the energy difference
 between the $1f_{5/2}$ and $2p_{3/2}$ states for the HO potential with standard
parametrization \cite{Sie87}. Additionally, during the minimization procedure
the energy difference  $e_{1p_{3/2}}-e_{1p_{1/2}}$ has been kept constant and equal to $\Delta_{ls}(p)$.
The results are displayed
in Table~\ref{table2}.

\begin{figure}[h]
\begin{center}
\vspace{1cm}
   \leavevmode
   \epsfxsize  = 12cm
     \epsfysize = 10cm
    \epsffile{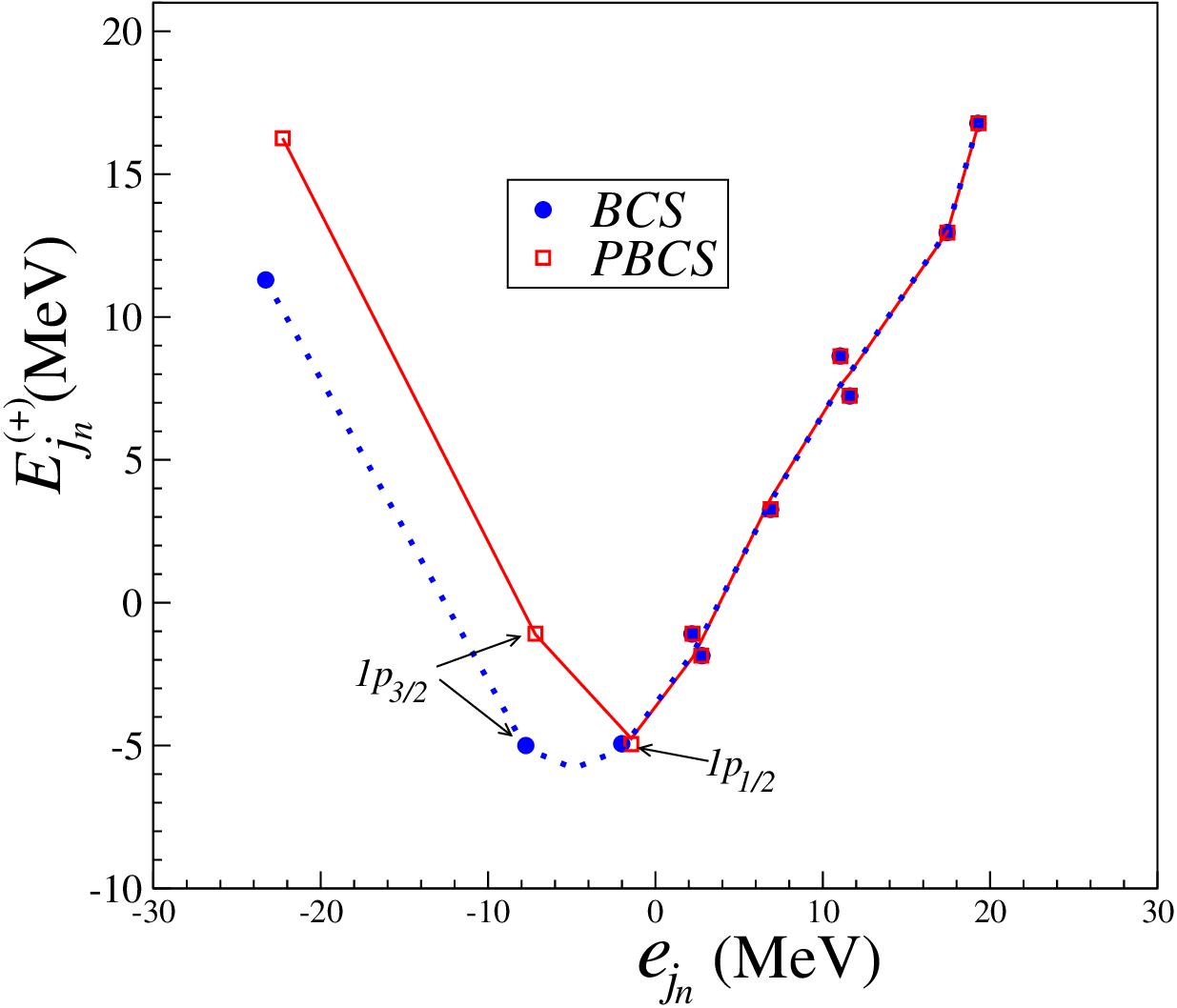}
\end{center}
\caption{\footnotesize Relationship between the single-particle
and the quasiparticle excitation energies $E_{j}^{(+)}$ for $^{13}$C.
The states
are ordered as $1s_{1/2}$, $1p_{3/2}$, $1p_{1/2}$, $2s_{1/2}$,
$1d_{3/2}$, $1f_{7/2}$, $2p_{3/2}$, $2p_{1/2}$, and $1f_{5/2}$,
and the energies are indicated by filled circles (BCS) and
unfilled squares (PBCS).} \label{fig1}\end{figure}
 The relationship between the s.p.e. $e_{j}$  and
the corresponding quasiparticle energies $E_{j}^{(+)}$ are illustrated in
Fig.~\ref{fig1}.
The major difference between the BCS and the PBCS
results  appears in
$E_{1s_{1/2}}^{(+)}$ and $E_{1p_{3/2}}^{(+)}$, which  have not
been used in the fitting procedure. Of course, opposite happens
with the quasihole energies $E_{j}^{(-)}$ (see Tables III and IV in
Ref. \cite{Krm05}).

\begin{figure}[h]
\begin{center}
\vspace{1.cm}
     \leavevmode
     \epsfxsize = 12cm
     \epsfysize = 10cm
     \epsffile{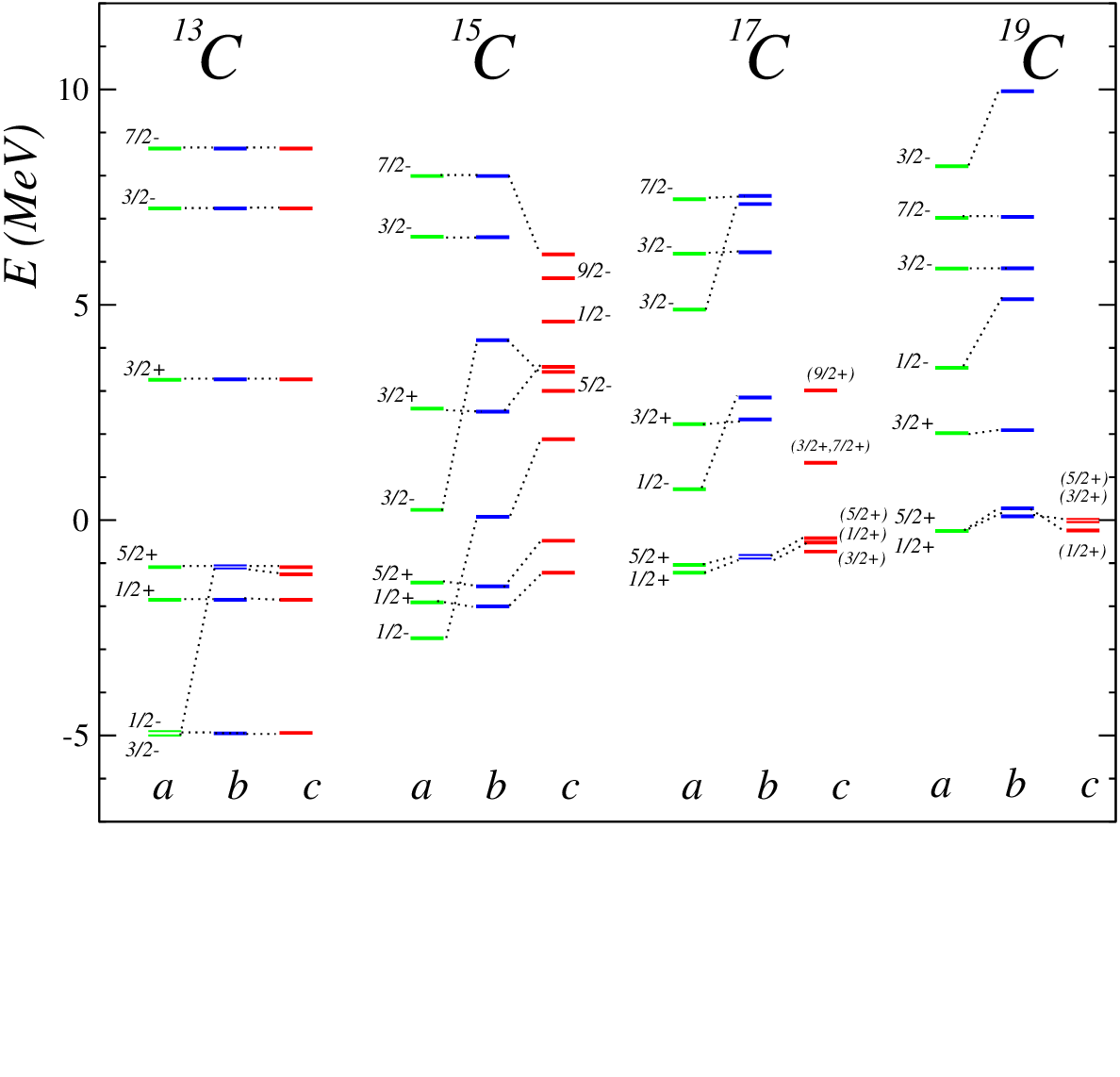}
   \end{center}
\caption{\small{Comparison between the calculated BCS, PBCS and
measured level  schemes for odd mass carbon isotopes: a) BCS, b)
PBCS, and c) experiments: from  Ref.~\cite{Ajz91} for $^{13}$C,
from  Refs.~\cite{Ajz91,Mur94,Boh03,Cap04} for $^{15}$C, from
Refs.~\cite{Ele05,Boh05} for $^{17}$C, and from  Ref.~\cite{Ele05} for
 $^{19}$C. For $^{13}$C are only shown the experimental levels that
 have been used in the fitting procedure.}}
\label{fig2}\end{figure}

\begin{figure}[t]
\begin{center}
\vspace{1.cm}
    \leavevmode
     \epsfxsize = 10cm
     \epsfysize = 11cm
     \epsffile{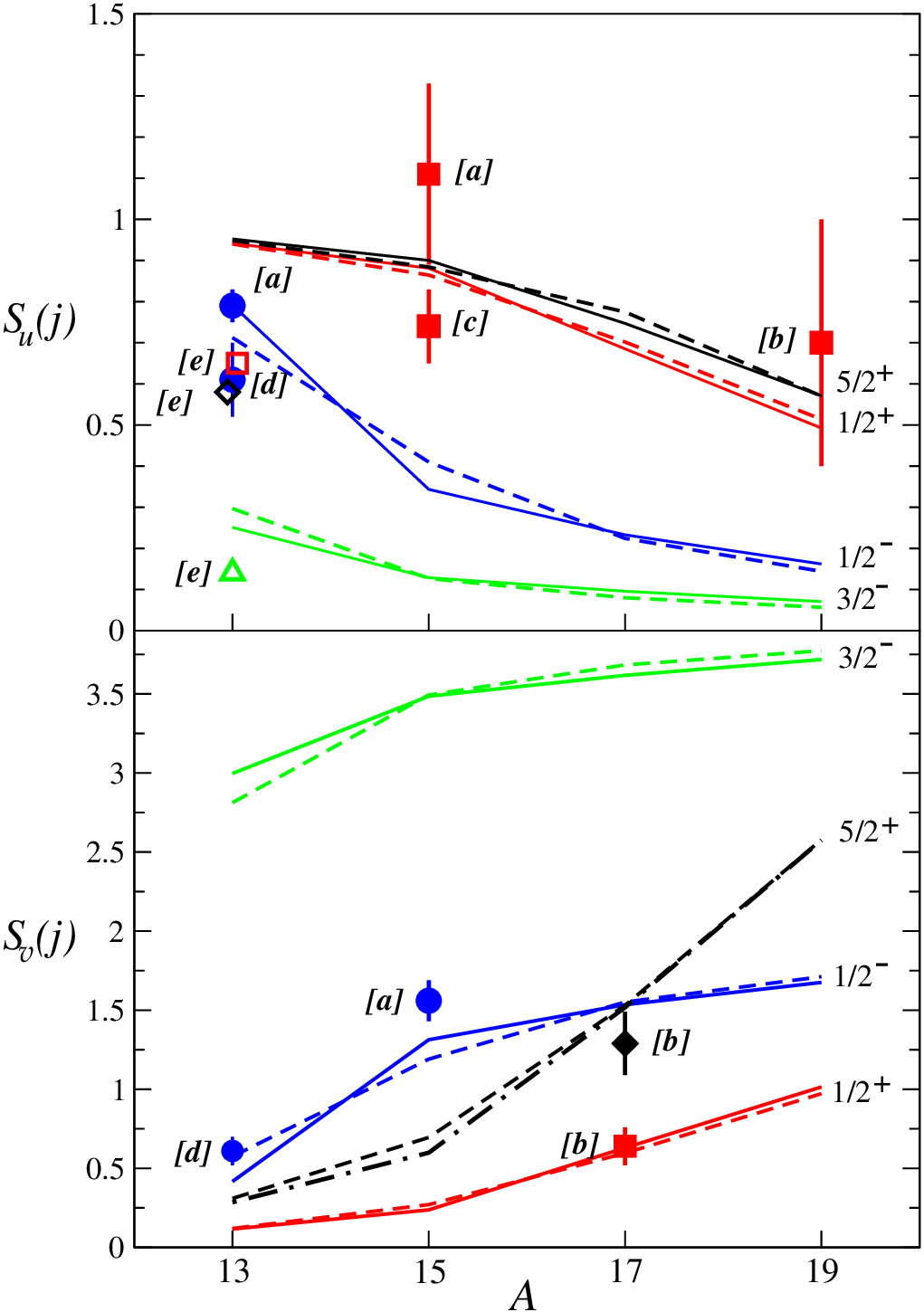}
   \end{center}
\caption{(Color online)
One-particle reaction spectroscopic factors as a function of the mass
number $A$. Upper panel:   $S_u(j)$,
for stripping on even parent and pick-up/ knock-out on odd parent.
Lower panel: $S_v(j)$, for pick-up/knock-out on even parent
and stripping  on odd parent. The
dashed and solid lines correspond, respectively, to the BCS and PBCS
predictions. The experimental values for the lowest
$J^\pi=1/2^-,3/2^-,1/2^+$ and $5/2^+$ states are indicated, respectively,
with circles, triangles, squares, and diamonds. The data  were by
recompiled from Refs.:
\cite{Tsa05}$^{[a]}$, \cite{Mad01,Bro01}$^{[b]}$,\cite{Bro01,Baz98}$^{[c]}$,
\cite{Liu04}$^{[d]}$  and \cite{Ohn85}$^{[e]}$.}
\label{fig3}\end{figure}

After fixing the parameterizations ($e_j$ and  $v_{{s}}^{{pair}}$)
in $^{13}$C, we evaluate the low-lying energy spectra for the
remaining odd-mass carbon isotopes, by changing only the number
of neutrons. In other words, we solve the BCS and PBCS gap
equations with the parameters listed in Table~\ref{table2} and the
number of particles $N=8,10$, and $12$ for $^{15}$C, $^{17}$C and
$^{19}$C, respectively. The results for the energy spectra and the
spectroscopic factors are shown
 and compared with experimental data
~\cite{Ajz91,Mur94,Boh03,Cap04,Ele05,Boh05,
Tsa05,Mad01,Bro01,Baz98,Liu04,Ohn85}
in Figs.~\ref{fig2}
and~\ref{fig3}, respectively.
One immediately sees that  the pairing
interaction reproduces to a significant extent the systematic of
the energy spectra of these nuclei.
 In particular, the decrease  in the separation
energies in going from $^{13}$C to  $^{19}$C is fairly well
accounted for
\footnote{It would be very interesting to compare our
results with a  systematic large scale shell-model calculations
of  all four odd-mass carbon isotopes, employing the same
Hamiltonian and the same single-particle space.}.

 The experimentally observed spin ordering $1/2^+, 5/2^+, 1/2^-$
 of the lowest three states in  $^{15}$C, is well  reproduced within
 the PBCS, indicating that the number projection plays a
 significant role,  and that these levels could be predominantly
 one-quasiparticle (1qp) states. The present model also accounts
fairly well for
 the $3/2^-$ state at $4.66$ MeV and the $3/2^+$ state at $4.78$ MeV.
  However, the  first one of these two states could be quite likely the
  partner of the  $5/2^-_1$ state at $4.22$ MeV, possessing therefore
 large, if not  dominant, three-quasiparticle (3qp) components
$\ket{(1d_{5/2})^2,2;1p_{1/2}}$ and $\ket{(2s_{1/2})^2;2,1p_{1/2}}$,
as suggested by the work of Bohlen \etal~\cite{Boh03}.

In $^{15}$C a doublet is known from the two-neutron transfer
$^{13}$C(t,p)$^{15}$C   reaction \cite{Tru83} with possible spin
assignment of ($9/2^-$,$7/2^-$),  at almost the same
energies ($6.84$ MeV, $7.39$ MeV) as the $(3^-,2^-)$ doublet in
$^{14}$C.
Based on this fact, one can speculate that their 3qp
structure is dominated by the  $\ket{(1d_{5/2})^2,4;1p_{1/2}}$
 configuration~\cite{Boh03}.
The  $1/2^-$ level  at $5.87$ MeV could be a 3qp
(seniority one or three) state which is not contained in our
configuration space. In fact, our restriction to 1qp subspace
eliminates many of the well known states in $^{15}$C.
However, except for the $5/2^-_1$ state, the  energy spectrum compares
fairly well with
the experimental results and with a recent shell-model
calculation (cf. Fig. 2 in Ref.~\cite{Suz03}).

The inversion of the $5/2^+_1$ and $1/2^+_1$ states  in $^{15}$C,
regarding the standard  single-particle ordering of levels, which
occurs  in $^{17}$O, has been discussed a long time ago by  Talmi
and Unna~\cite {Tal60}. They   have shown, in a brilliant manner,
that the crossing of these two levels, in going from $^{17}$O to
$^{15}$C, comes from difference in interaction energy of the
$1d_{5/2}$ and  $2s_{1/2}$ neutrons with two removed $1p_{1/2}$
protons
\footnote{The fact that $^{17}$C has a low-lying $1/2^+_1$ state,
and $^{19}$C has a $1/2^+_1$ ground state could also be  a consequence
of the same $1p_{1/2}^{-1}1d_{5/2}$  and $1p_{1/2}^{-1}2s_{1/2}$ ($T=1$)
particle-hole matrix elements, lowering configurations involving the
$1/2^+_1$ states in $^{19}$O and $^{21}$O at $1.47$ MeV and $1.22$ MeV,
respectively, very much in the same way as for $^{15}$C.}.
Nevertheless, as this phenomenon  is typical of weakly-bound light
neutron rich nuclei, it is seldom considered to be an ``exotic''
feature of $^{15}$C~\cite{Cap04,Thi04}.
That is, it  is attributed to the halo formation, which makes the
lowest angular momentum to gain energy by extending its wave
function. This apparent ``anomaly''
even takes place in the non-exotic $^{13}$C nucleus, and in our model it can be
interpreted as
a consequence of the pairing interaction. In fact, as
seen from Table~\ref{table2} and Fig.~\ref{fig2}, we obtain that  for
all odd carbon isotopes is $E^{(+)}_{1d_{5/2}}> E^{(+)}_{2s_{1/2}}$, although
$e_{1d_{5/2}}< e_{2s_{1/2}}$.
 This is a direct  consequence of fact that the
$2s_{1/2}$ self-energy is significantly larger than that of the
$1d_{5/2}$ state, because of the strong interaction between the
$1s_{1/2}$ and $2s_{1/2}$ states (see Eq. \rf{2.2}).
It might be worthwhile to  point out
that the
 effective s.p.e. $e_j$ used here should not  be confounded
with the $^{17}$O energy spectrum.

The separation energy of the last neutron in  $^{17}$C is
$S_n=729\pm 18$ keV \cite{Til93},
and the shell model calculation \cite{War92} predicts a
$J^\pi=3/2^+$ ground state.
This prediction has been confirmed later on by
the single-neutron knockout reaction  measurements
done by Maddalena \etal  \cite{Mad01}, which  strongly  indicates
such an assignment instead of   the naively expected option  $J^\pi=5/2^+$
arising  from the seniority-one state
$\ket{(1d_{5/2})^3{J^\pi=5/2^+}}$.
A simple explanation for this experimental result could be found in
the so called $J=j-1$ anomaly discussed by Bohr and Mottelson
\cite{Boh75}. In fact, since the work of Kurath \cite{Kur50}  we know
that for $(j)^3$ configurations with $j\ge 5/2$, the $J=j-1$ state
can occur below the $J=j$ state for  sufficiently long range forces
\footnote{The $3/2^+$  ground state in $^{17}$C  can similarly be
attributed to the differential effect  of above mentioned particle-hole
interaction on the $2s_{1/2}$  content of the dominantly $(1d_{5/2})^3$
$5/2^+_1$  and $3/2^+_1$ states of $^{19}$O.}.
The model discussed here does not contain
seniority-three states and therefore it is unable to
account for the $3/2^+$ ground state spin.
However, as seen from the Fig.~\ref{fig2}, it predicts
that the first two excited states  are $1/2^+$ and $5/2^+$ (in this order),
which is consistent with the recent measurement
done by Elekes \etal \cite{Ele05}.

Special attention was given to the neutron-rich carbon isotope
$^{19}$C to establish whether it has  a pure $s_{1/2}$ one-neutron-halo
as suggested in Refs. \cite{Rid98,Baz95}. On the basis of
measurements of different observables, associated with the
neutron-removal reaction  \cite{Mad01,Nak99,Cor01}, there seems
to be the consensus that the spin and parity of its
ground state is $J^\pi=1/2^+$. Contrarily, there is a strong
discrepancy in the literature regarding the separation energy of
 $^{19}$C. The tabulated values go from $S_n=160\pm 95$ keV in 1993-1997
\cite{Aud93} to  $S_n=580\pm 90$ keV in 2003 \cite{Aud03}.
 The experiments using time-of-flight techniques suggest small
separation energy, that is, weighted average yields $S_n=242\pm 95$ keV
\cite{Baz95}.
 The Coulomb dissociation of $^{19}$C was studied by Nakamura
in \cite{Nak99}, and the analysis of angular distributions of breakup
products suggests the value $S_n=0.53\pm 0.13$ MeV. Using this value in
the simple  cluster model calculation of the dipole strength gives
good agreement with the data. On the other hand, more recent  experiment
of Maddalena {\it et al.}
\cite{Mad01} on nuclear breakup of $^{19}$C yields $S_n=0.65\pm 0.15$
MeV and $0.8\pm 0.3 $ MeV.
In the present calculation we correctly reproduce the spin and parity
of the ground
state. For its  energy we obtain  $-0.33$ MeV in the BCS and  $0.12$
MeV in the PBCS.  Both results are in fair agreement
with the values reported in Refs. \cite{Aud93,Baz95}.
Two excited states at energies of $197(6)$ keV and $269(8)$ keV were
reported in a recent study of the $\gamma$-ray spectra \cite{Ele05}.
 The suggested
spins and parities are $3/2^+$ and $5/2^+$, and, as seen from Fig.~\ref{fig2},
we reproduce  the second one only. The level $3/2^+$  is very likely
dominated by the seniority-three configuration
$\ket{(1d_{5/2})^3{J^\pi=3/2^+}}$, same as  the  ground state in  $^{17}$C.

The BCS and PBCS one-particle reaction spectroscopic factors for the
lowest $J^\pi=1/2^-,3/2^-,1/2^+$ and $5/2^+$ states in odd-mass
carbon isotopes are shown
in Fig.~\ref{fig3}, as a function of the mass number   $A$.
They are quite similar to each other,  and,
except for $^{13}$C, they also agree fairly well with
the experimental data  which are
displayed in the same figure.
This  agreement  clearly implies that in  $^{15}$C- $^{19}$C
nuclei the states $J^\pi=1/2^-,1/2^+$ and $5/2^+$
 are basically seniority-one states.
In contrast,  the contribution
of  seniority-three  states, and/or of the
collective degrees of freedom seems to be quite relevant in $^{13}$C.
The latter will be discussed in the next section.

\section{$^{13}$C  and $^{11}$Be within  the Quasiparticle-Core Model}

The quasiparticle-rotor model hamiltonian is evaluated in the same way as
in the PRM~\cite{Nun95,Nun96,Esb95,Rid98}, except for:\\
1. The band-head energies are
modified as:
\be
e_j\go
E_j^{(+)}
\label{3.12}
\ee
with $E_j^{(+)}$ defined in Table~\ref{table1},
 and\\
2. The non-diagonal particle-core matrix elements are renormalised
by the overlap factors:
\be
F^{QP}_{jj'}=u_ju_{j'}-v_jv_{j'}.
\label{3.13}\ee
in the  QPRM, and
\be
F^{PQP}_{jj'}=\frac{u_ju_{j'}I^N(jj')-v_jv_{j'}I^{N-2}(jj')}
{[I^N(j)I^{N}(j')]^{1/2}},
\label{3.14}\ee
in the PQPRM.
\begin{table}[h]
\caption{
Neutron Wood-Saxon s.p.e. $e_j$ for $^{13}$C and $^{11}$Be
and  the corresponding quasiparticle BCS and
PBCS energies $E^{(+)}_j$, together the energies
$E(2^+)$ of the collective $2^+$ states.  All energies are given in
units of MeV, and the  pairing strength is
$v_{{s}}^{{pair}}=30$.
}
\begin{center}
\label{table3}
\newcommand{\cc}[1]{\multicolumn{1}{c}{#1}}
\renewcommand{\tabcolsep}{.6pc}
\bigskip
\begin{tabular}{c| rrr| rrr}\hline \hline
$Shell$       & &$^{13}$C& & &$^{11}$Be& \\ \hline
       &$e_{j}$
&&$E^{(+)}_j$~~~~~~~~~~~~&$e_{j}$&&$E^{(+)}_j$~~~~~~~~~~~~\\
       &        &BCS&PBCS&            &BCS&PBCS\\
\hline
$1p_{3/2}$ &$-13.54$&$-9.5956$ &$-7.109$ &$-7.33$&$-3.5928$&$-1.8592$\\
$1p_{1/2}$ &$-7.82$ &$-10.4057$ &$-12.209$&$-2.59$&$-3.9583$&$-5.4577$\\
$1d_{5/2}$ &$ -0.82$ &$ -4.2699$&$-4.123$ &$3.50$ &$ 0.2640$&$0.2333$\\
$2s_{1/2}$ &$-0.61$ &$-3.4033$ &$-3.303$ &$0.30$ &$-2.0479$&$-2.1316$
\\ \hline
$E(2^+)$      &      &&$4.438$&&&$3.368$ \\
\hline
\end{tabular}
\end{center}
\end{table}

Thus,  the PRM differs in several important aspects from the QPRM
and  PQPRM.
First, as shown in Fig.~ \ref{fig1}, the BCS and PBCS energies
$E_j^{(+)}$ can be  quite different from the s.p.e. $e_j$, not
only in magnitude but also in sign.
Second, the factors $F_{jj'}$, which are quite similar in BCS and
PBCS, correctly take in account the PP
and they  can by considerable less than unity for states near
the Fermi level, diminishing  in this way their coupling to the core
excited states quite a lot. In addition,
the particle-like states do not couple to the hole-like states, and
if the particle-core coupling is attractive (repulsive) for a
particle-like state $j$, it is repulsive (attractive) for a hole-like
state with same quantum numbers $j$.

\begin{figure}[t]
\begin{center}
\vspace{1.cm}
    \leavevmode
     \epsfxsize = 10cm
     \epsfysize = 11cm
     \epsffile{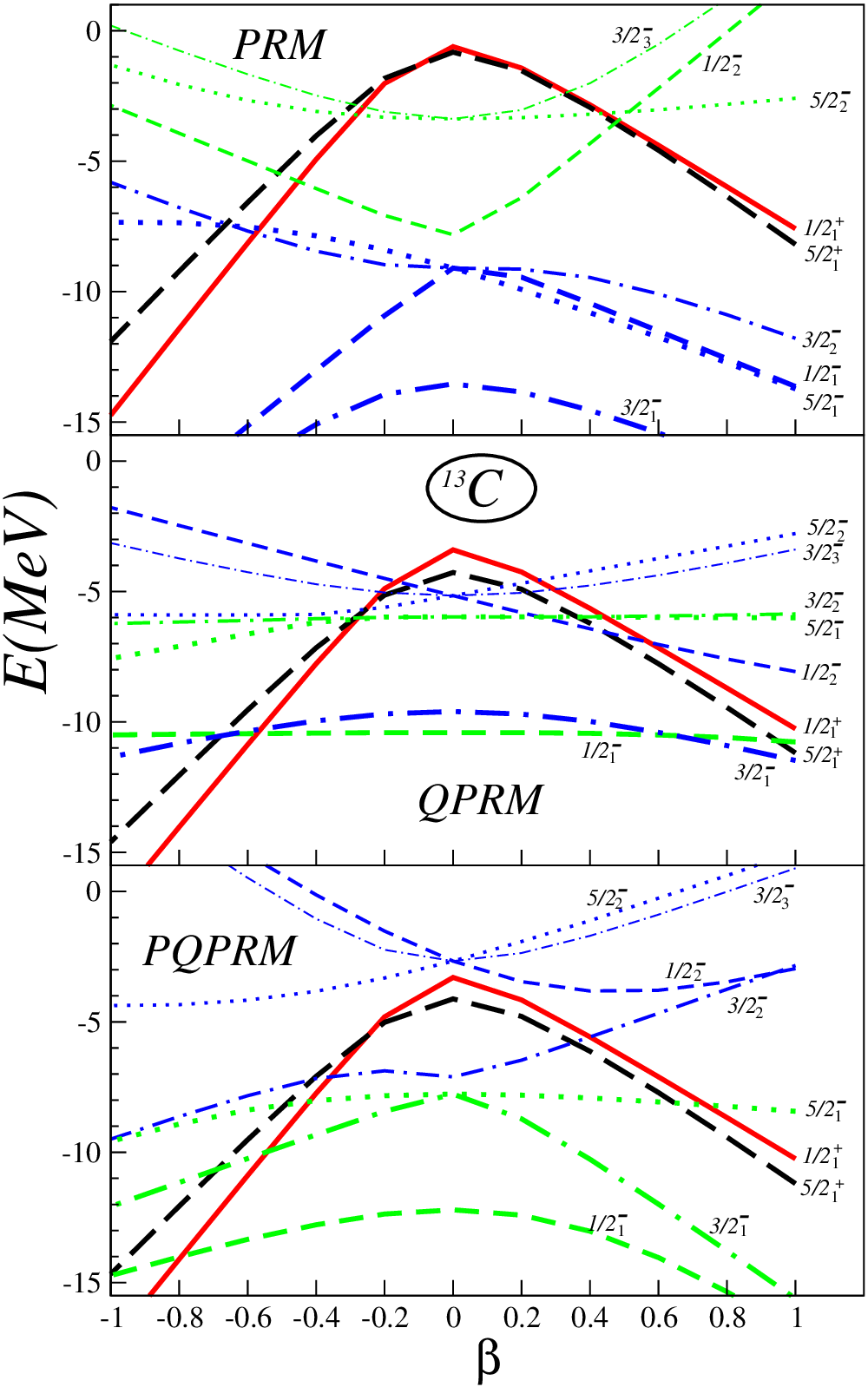}
   \end{center}
\caption{(Color online)
Calculated levels for   $^{13}$C as functions of the $\beta$
value for  $^{12}$C, within the PRM, QPRM and PQPRM models.}
\label{fig4}\end{figure}


The vibrational analog of the QPRM, \ie the quasiparticle-vibrator
coupling model (QPVM) has been introduced in Refs. \cite{Ste71,Bar75,Boh75}.
Note also that when  only the collective state $2^+$
is considered, and the  diagonal $2^+-2^+$ interactions are neglected,
the QPRM and  QPVM formally yield the same result,
except that the negative value of the deformation parameter
$\beta$ does not have any physical
meaning in the vibrational case. The same statement is valid for the
projected version of the QPVM, \ie for the PQPVM.

Incidentally  the s.p.e., used in the previous section and
shown in Table~\ref{table2}, were extracted from
the experimental data and thus  already include
the collective degrees of freedom. As such, they
 can not be used in the core-particle models.
Moreover,
due to the fact that the results strongly depend on the size of the
configuration space, and in order to make the comparison with
previous particle-core calculations~
\cite{Vin95,Col01,Nun95,Nun96,Esb95,Rid98} as close as possible,
we will use here only four bare
 s.p.e., obtained from a Wood-Saxon potential with  standard
parametrization \cite{Boh69}.
The results for $^{13}$C and $^{11}$Be are shown in Table~\ref{table3}.
Being the single-particle space here  smaller than the one used in the
previous section, we employ now a somewhat larger pairing strength,
in order to achieve the convergence of the BCS equations. We wish to
stress  that all the discussion done within the particle-core models is
basically qualitative, and therefore the choice of the model parameters
(Wood-Saxon potential, $E(2^+)$, $v_{{s}}^{{pair}}$, and $\beta$) doesn't play
a crucial role here.

\begin{figure}[t]
\begin{center}
\vspace{1.cm}
    \leavevmode
     \epsfxsize = 12cm
     \epsfysize = 10cm
     \epsffile{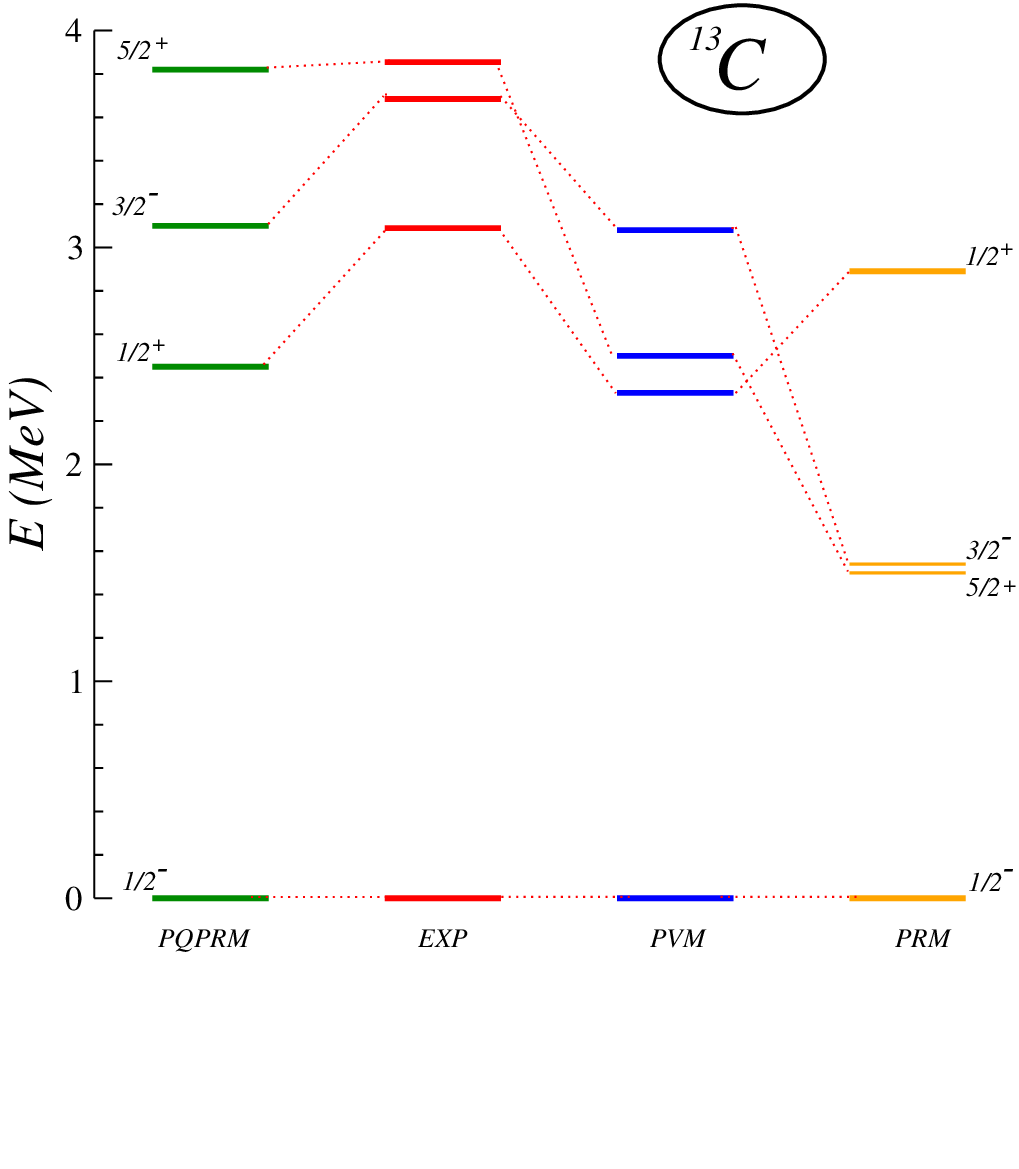}
   \end{center}
\vspace{-1.cm}
\caption{(Color online)
\small{ The low-energy spectra in  $^{13}$C within the  PQPRM
for $\beta=-0.6$, compared with the experimental levels (EXP)
and the PVM \cite{Vin95} and PRM \cite{Nun96} calculations.}}
\label{fig5}\end{figure}

The calculated low-lying levels for   $^{13}$C as functions of the $\beta$
value for  $^{12}$C, within the PRM, QPRM and PQPRM models are shown
in Fig.~\ref{fig4}.
Let us first note that
 the lowest states $J^\pi=3/2^-_1,1/2^-_1,3/2^-_2$ and
$5/2^-_2$, based on the unperturbed configurations
$\ket{1p_{3/2},0^+}$ and $\ket{1p_{3/2},2^+}$,
are frequently just ignored within the particle-core coupling models
\cite{Nun95,Nun96,Vin95,Col01} in order to simulate the PP
\footnote{ In Ref.~\cite{Esb95} have not been considered the negative
  parity states.}.
We are  considering them, however, in order to
make the comparison with the   QPRM and PQPRM models, where these states
are physically  meaningful.
Moreover,  the state
$J^\pi=1/2^-_2$, build up on
$\ket{1p_{1/2},0^+}$, is considered to be the ground state in the PRM.

The measured quadrupole moment of the
core $^{12}$C is $Q_0=-(22\pm10)$~e~fm$^2$ \cite{Ver83},
suggesting an oblate shape. This gives a quite large quadrupole
deformation ($\beta\cong -0.6$). As can be seen
from Fig.~\ref{fig4},  only the PQPRM reproduces
satisfactorily the experimental energy  ordering of the lowest four levels
in $^{13}$C with this value of $\beta$.
In Fig.~\ref{fig5} we confront the  PQPRM
 energies for this $\beta$ of the lowest $1/2^-$,
$3/2^-$, $1/2^+$ and $5/2^+$ levels  in  $^{13}$C with
the experimental data, and the PVM \cite{Vin95} and PRM
\cite{Nun96} calculations.

It should be noted that  the PRM and QPRM are unable to account for
the  $^{13}$C energy
spectra for  any value of the deformation parameter,
neither positive nor negative. In the first case, and even when
the levels based on the configurations
$\ket{1p_{3/2},0^+}$ and $\ket{1p_{3/2},2^+}$
are omitted, a low lying $5/2^-$ state, arising  from the unperturbed
$\ket{1p_{1/2},2^+}$ level, shows up, which is not observed experimentally.
 On the other hand, in the second case,
the particle-core coupling never removes the degeneracy
between the states  $1/2^-_1$ and
$3/2^-_1$.

The resulting PRM, QPRM and PQPRM wave functions for the lowest
$1/2^-$, $3/2^-$, $1/2^+$, and  $5/2^+$
states,  with $\beta=-0.6$,  are confronted with previous calculations
\cite{Vin95,Esb95,Nun96} in  Table~\ref{table4}.
As mentioned before, within  the PRM are ignored the
states $J^\pi=3/2^-_1,1/2^-_1$, and
$3/2^-_2$, based on the unperturbed configurations
$\ket{1p_{3/2},0^+}$ and $\ket{1p_{3/2},2^+}$, and are exhibited
the wave functions of  the $1/2^-_2$
and  $3/2^-_3$ levels.
On the other hand, in the PRM calculations done in Refs
\cite{Esb95,Nun96}
instead of the wave functions are presented
the percentages of the different configurations,
and therefore   we only show  absolute values
of the corresponding amplitudes.

It is worth noting that, while  the pairing makes the wave functions of
the negative parity states less collective,
all three models used here yield similar results for the
positive parity states.
We find that about $55 \% $ of the $1/2^+_1$ state consists
of the $1d_{5/2}$ single-particle state coupled to the $2^+$ excited
core state.
In contrast, in the previous particle-core coupling  calculations
\cite{Esb95,Vin95,Nun96}  this state is basically
($\ge 90 \%$) of single-particle nature.
As seen from the Table~\ref{table4},
our wave functions for  the  $5/2^+_1$ are also significantly more
collective
that those obtained in the just mentioned works.

\begin{table}[h]
\caption{\footnotesize Comparison between the wave functions and the
spectroscopic factors obtained in
the present work for the lowest states in $^{13}$C with those derived by
Vinh Mau \cite{Vin95} within the PVM,  and by  Nunes \etal \cite{Nun96}
and Esbensen \etal  \cite{Esb95} within the
PRM, and the shell model study done by   Cohen and  Kurath~\cite{Coh67}.
 In all our calculations (PRM, QPRM, and PQPRM) we adopt
$\beta=-0.60$.}
\begin{center}
\label{table4}
\newcommand{\cc}[1]{\multicolumn{1}{c}{#1}}
\renewcommand{\tabcolsep}{0.01pc} 
\renewcommand{\arraystretch}{0.5} 
\begin{tabular}{r| r| c r r r| c c c c}
\hline
\multicolumn{1}{c|}{$J^\pi$}&
\multicolumn{1}{c|}{Model}&
\multicolumn{4}{c|}{Wave Functions}&
\multicolumn{4}{c}{Spectroscopic Factors}
\\ \hline
&&~$\ket{1p_{1/2};0^+}$
&$\ket{1p_{1/2};2^+}$
&$\ket{1p_{3/2};0^+}$
&~$\ket{1p_{3/2};2^+}$~&~~Theor.&&Exp.&
\\
&&&&&&&~Ref.\cite{Pea72}~&~Ref.\cite{Ohn85}~&~Ref.\cite{Liu04}
\\ \hline
$1/2^-$
&PRM      &$0.849$  &&&$0.528$~~&$0.72$&&&
\\
&QPRM     &$0.997$  &&&$0.076$~~&$0.69$&&&
\\
&PQPRM    &$0.961$  &&&$0.277$~~&$0.53$&$0.58~(15)$&$0.77$&$0.61~(9)$
\\
&~PVM~\cite{Vin95}~&$0.791$&&&$ 0.602$~~&$0.63$&&&
\\
&~PRM~\cite{Nun96}~&$|0.565|$&&& $|0.818|$~~&$0.32$&&&
\\
&SM~\cite{Coh67}~&&&& &$0.61$&&&
\\\hline
$3/2^-$
&PRM          &&$0.929$  &$-0.162$  &$0.332$~~&$0.03$&&&
\\
&QPRM         &&$-0.052$&$0.935 $ & $ 0.350$~~&$0.17$&&&
\\
&PQPRM        &&$0.759$&$-0.650$&$0.025$~~&$0.10$&&$0.14$&
\\ 
&~PRM~\cite{Nun96}~&&$|0.787|$ &$| 0.375|$&$|0.489|$~~&$0.14$&&&
\\
&SM~\cite{Coh67}~&&&& &$0.19$&&&
\\\hline \hline
&
&~$\ket{2s_{1/2};0^+}$&$\ket{2s_{1/2};2^+}$
&~$\ket{1d_{5/2};0^+}$&$\ket{1d_{5/2};2^+}$~&&&&
\\ \hline
$1/2^+$  &PRM   &$0.684$&&&$0.730$~~&$0.47$&&&
\\
         &QPRM  &$0.672$&&&$0.741$~~&$0.44$&&&
\\
         &PQPRM &$0.671$&&&$0.741$~~&$0.44$&$0.36(2)$&$0.65$&
\\
&~PVM~\cite{Vin95}~&$ 0.957$&&&$0.291$~~&$0.92$&&&
\\
&~PRM~\cite{Esb95}~&$|0.975|$&&&$|0.200|$~~&$0.95$&&&
\\
&~PRM~\cite{Nun96}~&$|0.952|$&&&$|0.272|$~~&$0.90$&&&
\\
&~SM~ \cite{Mil89}~& $0.945$ &&&$0.251$~~&$0.89$&&&
\\
\hline
$5/2^+$   &PRM  &&$0.485$&$0.737 $   &$-0.471$~~&$0.54$&&&
\\
          &QPRM &&$0.463$&$0.752 $   &$-0.469$~~&$0.55$&&&
\\
          &PQPRM&&$0.466$&$0.749$    &$-0.471$~~&$0.55$&&$0.58$&
\\
&~PVM~\cite{Vin95}~&& $0.179$&$ 0.867$ &$0.300$~~&$0.75$&&&
\\
&~PRM~\cite{Esb95}~&& $|0.144|$&$| 0.894|$ &$|0.424|$~~&$0.80$&&&
\\
&~PRM~\cite{Nun96}~&& $|0.565|$& $|0.831|$ &$| 0.538|$~~&$0.69$&&&
\\
&~SM~ \cite{Mil89}~&& $?$& $0.897$ &$-0.357$~~&$0.80$&&&
\\
\hline\hline
\end{tabular}
\end{center}
\end{table}

The differences in the wave functions, displayed in the
Table \ref{table4}, are reflected in the
 reaction spectroscopic factors for pickup  on the $^{13}$C
target and for stripping on the  $^{12}$C target, which are
shown in the same table.
Also are listed  the  results of the previous PVM and
PRM~\cite{Vin95,Nun96,Esb95}, and  shell model~\cite{Coh67}
studies,
as well as the experimental results~\cite{Pea72,Ohn85,Liu04},
which are accounted for quite well
within the present PQPRM calculations.

We obtain
much smaller amplitudes for the non-collective components $
\ket{2s_{1/2};0^+}$ and  $\ket{1d_{5/2};0^+}$ of
the lowest  $\ket{\rm 1/2^+_1}$ and $\ket{\rm 5/2^+_1}$
states than do the full $1\hbar\omega$ shell-model
calculations, among which   the work  of
 Jager \etal \cite{Jag71}  was very   likely  the first one.
There have been  many
such calculations since then with refined effective interactions, but the
wave functions for the lowest positive-parity states remain essentially
unchanged. We show the ones obtained in the work done by
Millener \etal \cite{Mil89},
which, as seen from  Table \ref{table4},
 yield  spectroscopic factors that are $\sim 30-60\%$
larger than the measured ones.

\begin{figure}[t]
\begin{center}
\vspace{1.cm}
    \leavevmode
     \epsfxsize = 10cm
     \epsfysize = 14cm
     \epsffile{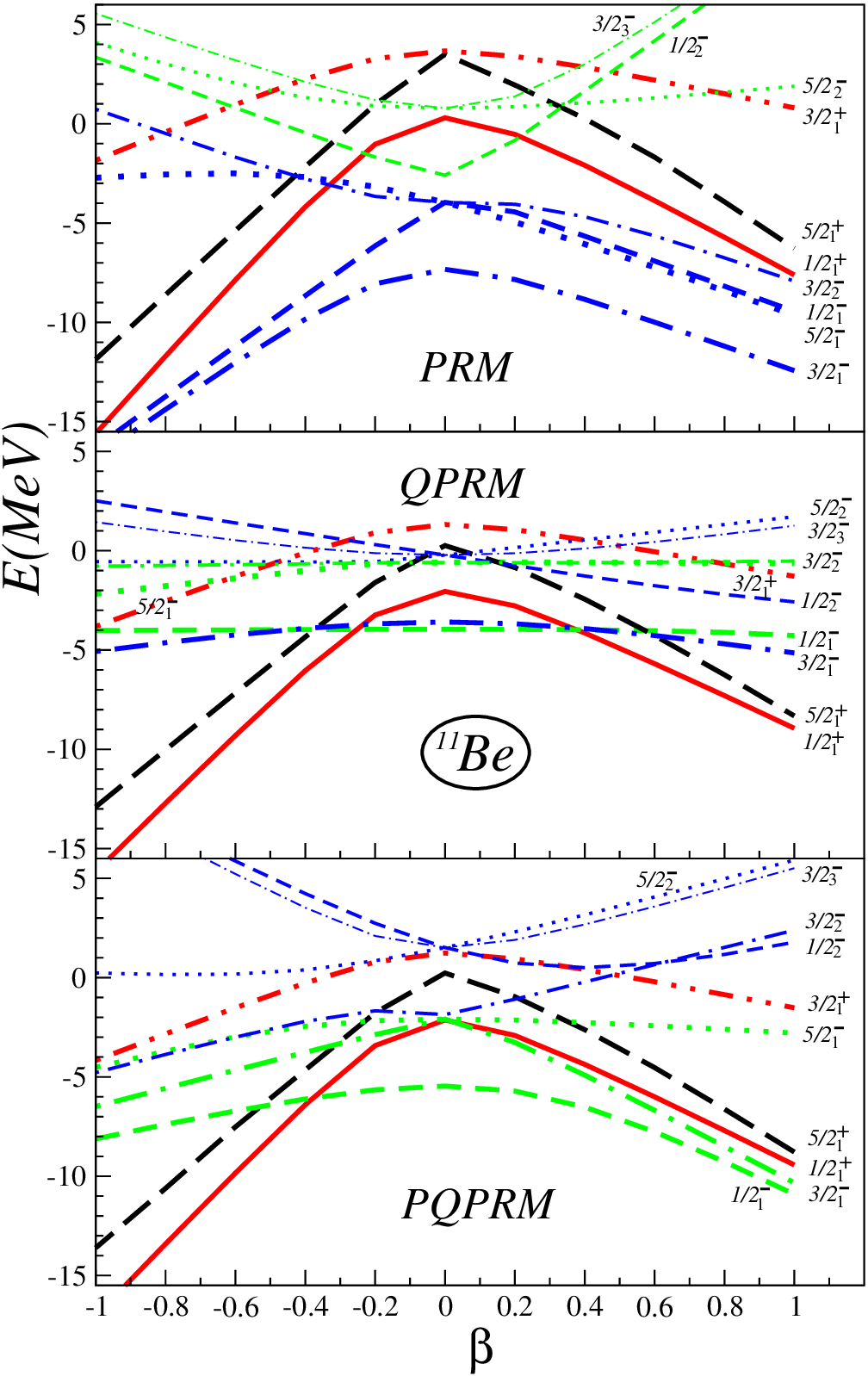}
   \end{center}
\caption{(Color online)
Calculated levels for $^{11}$Be as functions of the $\beta$
value for  $^{10}$Be, within the PRM, QPRM and PQPRM models.}
\label{fig6}\end{figure}

In Fig.~\ref{fig6} we show the PRM, QPRM and PQPRM results of our
study  of the low-lying
states in $^{11}Be$ as a function of deformation. All
what has been pointed out  in commenting the $^{13}$C nucleus in the PRM,
 regarding the PP, the unperturbed configurations
$\ket{1p_{3/2},0^+}$ and $\ket{1p_{3/2},2^+}$, and the negative parity
states, is also pertinent here.
On the other hand, when comparing our PRM results
with the work of Esbensen \etal \cite{Esb95} one sees
that their deformation dependence  of the levels $1/2^+$ and
$5/2^+$  gets flatter than ours for strong deformations. The reason for that
 is simple and comes from the fact that they   use a volume factor
in the calculation, given by \cite[(3)]{Esb95},
 to preserve the volume of the nucleus.
This actually means that their interaction radius is smaller for
higher deformations and this results in the flattening of the
curves.
Having said that, it should be noted that our PRM curves
for the states   $1/2_1^+$ and $1/2^-_2$ are very much like those
in  \cite[Fig. 2]{Nun96}. The main difference is that in our case the
crossings between  the positive and negative  $1/2$ states occur at
a significantly smaller value of  $\beta$.

Both  PRM calculations performed so far \cite{Esb95,Nun96} were done
with a positive value of $\beta$,
\ie a prolate deformation has been assumed  for the  $^{10}$Be
nucleus. However, there is no firm experimental evidence that it is so.
Furthermore, we neither know whether $^{10}$Be behaves as a rotator or
as a vibrator. From its energy spectra, with the first
and second $2^+$  at energies   $E_{2^+_1}=3.37$ MeV
 and $E_{2^+_2}=5.96$ MeV (\ie $E_{2^+_2}\cong 2 E_{2^+_1} $),
one can conclude that it is more likely a  vibrator. In fact,
Esbensen \etal  \cite{Esb95} have noted that for  the positive parity
spectra in  $^{11}$Be a better agreement with data
is obtained after reducing (in  $34\%$) the $2^+-2^+$
coupling strength. On the other hand,
Vinh Mau \cite{Vin95} and
Col\'o \etal \cite{Col01} use straightforwardly the PVM.

The most recent study of  quadrupole deformation of  $^{10}$Be has
been done through proton  inelastic scattering
\cite{Iwa00a}, and
the  values for  $\beta$,  extracted from
the measured deformation lengths $\delta=\beta R=\beta r_0
A^{1/3}$, are: $\beta=0.593(56)$
and $\beta=0.692(65)$.
A careful scrutiny of the energy levels in
 Fig.~\ref{fig6}, leads one to conclude
that none of our particle-rotor coupling
calculations is able to reproduce the experimentally observed
spin sequence  $1/2^+-1/2^--5/2^+$ in  $^{11}Be$,
for such a value of $\beta$  neither positive nor negative.
In view of this it does not make much sense to
comment the wave functions within the PRM,
QPRM and PQPRM approximations.

\begin{figure}[t]
\vspace{1.cm}
\begin{center}
    \leavevmode
     \epsfxsize = 10cm
     \epsfysize = 14cm
     \epsffile{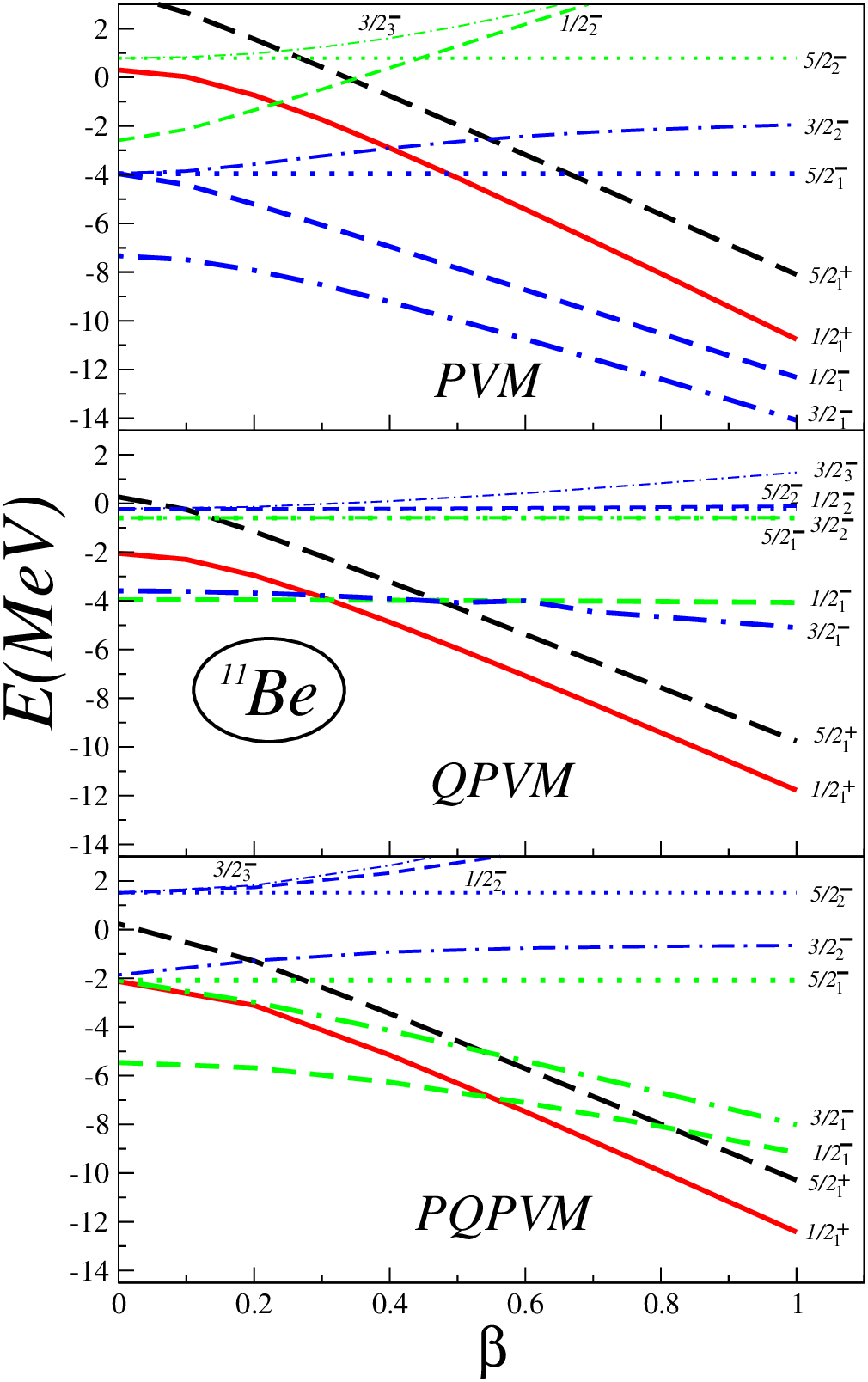}
   \end{center}
\caption{(Color online)
Calculated levels for   $^{11}$Be as functions of the $\beta$
value for  $^{10}$Be, within the PVM, QPVM and PQPVM models.}
\label{fig7}\end{figure}

The just mentioned results have induced us to replace  the rotor
by a harmonic vibrator in the description of the
low-lying   $^{11}$Be levels.
 The results are shown in Fig.~\ref{fig7}.
 The PVM (upper panel) exhibits  the same
features as the PRM regarding the PP. Therefore
 the lowest four negative parity states, same as in the PRM,
have to be discarded within the PVM.
However, even doing so
it is not possible to obtain the desired theoretical results.
On the other hand, from the QPVM energy spectra (middle panel) one sees  that
the particle-vibrator  coupling, same as the  particle-rotor coupling,
is unable to remove the degeneracy
between the states  $1/2^-_1$ and $3/2^-_1$.
Once more,  as seen from the PQPVM results (lower panel),
 this is  achieved through the number projection procedure only.

Both the  $1/2^+_1$ and $1/2^-_1$ levels
go down within the PQPVM  when $\beta$ is increased, and
their crossing happens close to the experimental value
for the deformation parameter ($\b\cong 0.6$).
Thus, the spin inversion mechanism is quite different here
than in the PRM and PVM cases, where the  $1/2^-_1$ state is pushed up
while the $1/2^+_1$ state is pushed down \cite{Nun96,Vin95}.
The fact that in the PQPVM the $1/2^-_1$ state is almost independent on the
value of $\beta$  while the  $1/2^+_1$ state varies very strongly is due to
the PP  factor given in the equation \rf{3.14}, which is small in the first case
($u_{p_{1/2}} \cong v_{p_{1/2}}$)  and large in  second case
($u_{p_{1/2}} \cong 1, v_{p_{1/2}}\cong 0$).
 Therefore, it can be
argued that within the particle-core model
the spin inversion in  $^{11}$Be is due to the PP.

\begin{figure}[t]
\begin{center}
\vspace{0.5cm}
    \leavevmode
     \epsfxsize = 12cm
     \epsfysize = 10cm
     \epsffile{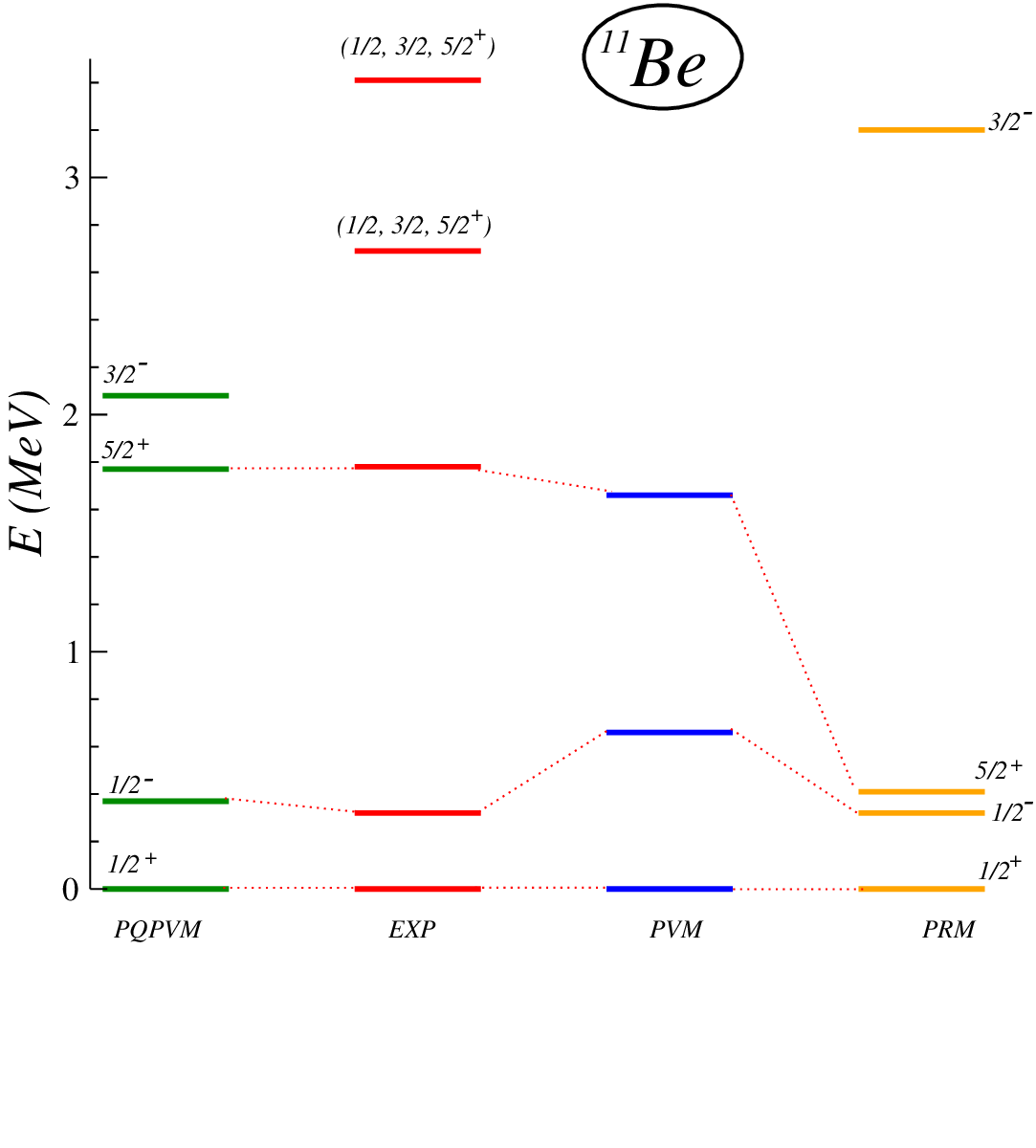}
   \end{center}
\caption{(Color online)
The low-energy spectra in  $^{11}$Be within the  PQPVM
for $\beta=0.6$, compared with the experimental levels (EXP)
and the PVM \cite{Vin95} and PRM \cite{Nun96} calculations.}
\label{fig8}\end{figure}

As shown in Fig.~\ref{fig8} all three lowest states in $^{11}Be$
are satisfactorily reproduced within the PQPVM for $\b=0.6$.
For the sake of  comparison in the same figure are also
presented the results of
 the PVM \cite{Vin95} and
PRM \cite{Nun96} calculations.


\begin{table}[h]
\caption{\footnotesize Comparison between the wave functions and the
spectroscopic factors obtained in
the present work for the lowest states in  $^{11}$Be with those derived by
Vinh Mau \cite{Vin95} within the PVM,  and by  Nunes \etal \cite{Nun96}
and Esbensen \etal  \cite{Esb95} within the
PRM, the  variation shell model (VSM) of Otsuka \etal \cite{Ots93}.
and the shell model study done by   Cohen and  Kurath~\cite{Coh67}.
 In all our calculations (PVM, QPVM, PQPVM) we adopt
$\beta=0.60$.}
\begin{center}
\label{table5}
\newcommand{\cc}[1]{\multicolumn{1}{c}{#1}}
\renewcommand{\tabcolsep}{0.01pc} 
\renewcommand{\arraystretch}{0.5} 
\begin{tabular}{ r| r| c r r r | c c c c}
\hline
\multicolumn{1}{c|}{$J^\pi$}&
\multicolumn{1}{c|}{Model}&
\multicolumn{4}{c|}{Wave Functions}&
\multicolumn{4}{c}{Spectroscopic Factors}
\\ \hline
&&~$\ket{1p_{1/2};0^+}$
&$\ket{1p_{1/2};2^+}$
&$\ket{1p_{3/2};0^+}$
&~$\ket{1p_{3/2};2^+}$~&~~Theor.&&Exp.&
\\
&&&&&&&~Ref.\cite{Aut70}~&~Ref.\cite{Zwi79}~&~Ref.\cite{Nav00}
\\ \hline
$1/2^-$ &PVM&$0.750 $&&&$-0.631$~~&$0.56$&&&
\\
&QPVM&$0.995$&&& $-0.103$~~&$0.63$&&&
\\
&PQPVM& $0.916$& &&$-0.401$~~&$0.49$&$0.63~(15)$&$0.96$&$0.45~(12)$
\\
&~PVM~\cite{Vin95}~&$0.746$&&&$0.667$~~&$0.56$&&&
\\
&~PRM~\cite{Nun96}~&$|0.932|$&&&$|0.362|$&$0.87$&&&
\\
&~SM~\cite{Coh67}~&&&&&$0.60$&&&
\\\hline 
$3/2^-$ &PVM&&$0.870$&$0.428$ &$-0.247$~~&$0.18$&&&
\\
&QPVM&&$0.070 $&$0.927 $&$-0.368$~~&$0.25$&&&
\\
&PQPVM&&$0.592$ &$0.735$ &$-0.331$~~&$0.15$&&&
\\
&~SM~\cite{Coh67}~&&&&&$0.17$&&&
\\\hline\hline
&
&~$\ket{2s_{1/2};0^+}$&$\ket{2s_{1/2};2^+}$
&$\ket{1d_{5/2};0^+}$&~$\ket{1d_{5/2};2^+}$~&&&&
\\ \hline
$1/2^+$
&PVM   &$0.826$  &&&$-0.564$~~&$0.68$&&&
\\
&QPVM  &$0.825$  &&&$-0.565$~~&$0.62$&&&
\\
&PQPVM &$0.821$  &&&$-0.571$~~&$0.64$&$0.73~(6)$&$0.77$&$0.53~(13)$
\\
&~PVM~\cite{Vin95}~& $0.964$&&&$0.267$~~&$0.93$&&&
\\
&~PRM~\cite{Esb95}~&$|0.933|$&&&$|0.317|$~~&$0.87$&&&
\\
&~PRM~\cite{Nun96}~&$|0.883|$&&&$|0.447|$~~&$0.78$&&&
\\
&~VSM~\cite{Ots93}~&$0.740$  &&& $0.630$~~&$0.55$&&&
\\\hline 
$5/2^+$ &PVM&   &$-0.523$&$0.740$&$0.422$~~&$0.54$&&&
\\
       &QPVM&   &$-0.486$&$0.760$&$0.375$~~&$0.56$&&&
\\
       &PQPVM & &$-0.485$&$0.758$&$0.437$~~&$0.56$&&$0.50$&
\\ 
&~PVM~\cite{Vin95}~&&$ 0.269$&$ 0.896$& $0.353$~~&$0.80$&&&
\\\hline \hline
\end{tabular}
\end{center}
\end{table}

The resulting PVM, QPVM and PQPVM wave functions for the lowest
$1/2^-$, $3/2^-$, $1/2^+$, and  $5/2^+$
states in $^{11}$B,  with $\beta=0.6$,  are confronted with
previous calculations
\cite{Vin95,Esb95,Nun96,Ots93} in  Table~\ref{table5}.
 All said above  in relation of omission of
 levels based on  configurations
$\ket{1p_{3/2},0^+}$ and $\ket{1p_{3/2},2^+}$,
 and the percentages of the wave function amplitudes  in
the case of $^{13}$C is valid here.

The PVM wave function for the $1/2^-_1$ state, obtained by
Vinh Mau ~\cite{Vin95}, is quite similar to the one we get within
the same model. Notice that the pairing tends to make this state
to be less collective.
The effect is still more pronounced for the $3/2^-_1$ state.

All three models  yields
similar results for  the positive parity   wave functions.
When  confronted with that
 derived  within the PVM by Vinh Mau \cite{Vin95}, one sees
that, at variance with what happens with negative  parity
states, our wave functions  are more
collective.
We also note that  our wave function for the $1/2^+_1$ state is
similar to that obtained in the variational shell model (VSM) 
calculation of Otsuka \etal Ref.~\cite{Ots93}.

The reaction spectroscopic factors for pickup  on the $^{11}$Be
target and for stripping on the  $^{10}$Be target,
 evaluated with the wave functions  listed in  Table~\ref{table5}
are presented in the same table.
We can read the differences with
the previous PRM results for the $1/2^+_1$ state~\cite{Nun96,Esb95},
and the similarity with the shell model calculation for the negative
parity states~\cite{Coh67}.It can be observed that
the experimental data
~\cite{Aut70,Zwi79,Nav00} for the levels $1/2^+_1,1/2^-_1$ and
$5/2^+_1$ are reproduced quite well by the PQPVM.

Before ending this section, and in order to inquire on how realistic
the above particle-core wave
functions are, we  briefly discuss the ground state magnetic
dipoles moments in $^{13}$C and $^{11}$Be. After some
Racah algebra and by denoting  $a$ and $b$, respectively, the
single-particle and collective components in the wave functions, we get:
\br
\mu(1/2^-_1)&=&
a^2\mu(1p_{1/2})-\frac{b^2}{3} \mu(1p_{3/2})+\frac{b^2}{2} \mu(2^+)
\nn\\
&=&-\frac{g_s}{6}+b^2g_R,
\label{3.30}\er
and
\br
\mu(1/2^+_1)&=&
a^2\mu(2s_{1/2})+\frac{7b^2}{15} \mu(1d_{5/2})-\frac{b^2}{3} \mu(2^+)
\nn\\
&=&\frac{g_s}{2}\left(a^2+\frac{7b^2}{15}\right)-\frac{2b^2}{3} g_R,
\label{3.31}\er
where $g_s=-3.82$ and $g_R=Z/A$ are, respectively,
 the effective spin and collective gyromagnetic ratios
 \cite{Boh69,Boh75}.

\begin{table}[h]
\caption{
Magnetic dipole  moments of the ground states in $^{13}$C ($1/2^-_1$) and
$^{11}$Be
 ($1/2^+_1$) evaluated with the wave functions listed in
 Tables~\ref{table4} and ~\ref{table5} are  confronted with the
 experimental data. PCM stands here for the particle-core model, with
 the core being a rotor in the case of $^{13}$C, and a vibrator in the
case of $^{11}$Be, etc.}
\label{table6}
\newcommand{\cc}[1]{\multicolumn{1}{c}{#1}}
\renewcommand{\tabcolsep}{0.12 pc}
\bigskip
\begin{tabular}{c|c c c c c  c c| c }
\hline \hline
Nucleus  && && Theory&&&& Experiment   \\ \hline
 & PCM & QPCM & PQPCM & Ref.\cite{Vin95}& Ref.\cite{Nun96}
& Ref.\cite{Esb95}
& Ref.\cite{Ots93} &Ref. \cite{Sto05}
\\ \hline
 $^{13}$C &$0.77$&$0.63$& $0.67$&$0.80$&$0.94$&$-$&$-$&$0.7024118(14)$
\\
$^{11}$Be&$-1.66$&$-1.67$& $-1.66$&$-1.85$&$-1.78$&$-1.71$&$-1.49$&$
-1.6816(8)$
\\ \hline
\end{tabular}
\end{table}

The numerical results are shown in  Table~\ref{table6}, from where one
can see that they the magnetic dipole moments
strongly depend on the interplay between the single-particle
and collective degrees of freedom.

\section{Summary and Conclusions}

Our work was divided into two main stages.
First, we apply pure BCS and PBCS models to describe heavy carbon isotopes.
We adopt the single-particle energies and pairing strengths which
reproduce simultaneously  the experimental  binding energies of
$^{11}$C and $^{13}$C,  and
the low-energy spectra of  $^{13}$C, except for
the $3/2^-_1$ state, which  is nicely reproduced
within the PBCS but not within the BCS (see Fig.~\ref{fig2}).
 With these parameters
we next evaluate  the low-lying states in $^{15}$C, $^{17}$C and
$^{19}$C. We found
 that  both models are capable to
 explain  fairly well  the decrease of the binding energies in going
from  $^{13}$C to $^{19}$C.
The PBCS model reproduces as well the experimentally observed spin ordering
$1/2^+,5/2^+,3/2^+$ of the lowest three states and the
energy of the first $3/2^-$ state in $^{15}$C. In addition, the same model
correctly predicts the lowest $1/2^+$ and $5/2^+$ levels in  $^{17}$C
and $^{19}$C, but it is not able to account for   the  $3/2^+_1$ level
in these two nuclei, which is very likely
 build up mainly on the anomalous seniority-three state
$\ket{(1d_{5/2})^3{J^\pi=3/2^+}}$.

We have calculated the one-particle transfer spectroscopic factors for the
lowest $1/2^-,3/2^-,1/2^+$ and $5/2^+$ states for all four odd-mass C
isotopes obtaining quite similar results within the two pairing models.
In  $^{13}$C the  calculations agree with the experimental
data for the
ground state, but they  fail to reproduce the experimental spectroscopic
factors for excited  states in this nucleus.
For the heavier carbon isotopes the
agreement with the experimental data is better. This implies that the
low-lying states in $^{15,17,19}$C are basically seniority-one states,
while the contribution of seniority-three states and the collective effects
could be important in $^{13}$C.
Our results clearly indicate that the pairing interaction plays a major
role in the nuclear structure
of heavy carbon isotopes, partly accounting for their basic feature such as:
a)  small binding energies, b) spin-parity
ordering  of the low-lying states, and
c) systematic decrease in the binding (of the order of $5$ MeV) when one goes  from $^{13}$C to
$^{19}$C. All this is achieved without employing any free parameter.
 Therefore, the pairing has to be taken into account in any theoretical
calculation that aspire to be quantitatively realistic.

In  the following  stage we  included the collective degrees of
freedom in the framework of the weak-coupling model
and applied it to describe
$^{13}$C and $^{11}$Be. Our main objective here was to analyze how the
short range pairing correlations modify the core-particle coupling
mechanism, and consequently the energy spectra, spectroscopic factors,
and the magnetic dipole moments  in these nuclei.
As far as we know such a study has not been
done so far, at least not in a systematic way
\footnote{See however the Refs. ~\cite{Gor04,Bro04,Bar01}.}
. It is worthwhile to stress
once more that the PP, which is
usually omitted in the simple  particle-rotor and particle-vibrator
coupling models, is  brought up back   into the play by the inclusion
of the pairing.
Here the single-particle energies were taken from the
standard parametrization for the
Wood-Saxon potential, while
the coupling matrix elements were calculated using
the wave functions of the harmonic oscillator. The last procedure
could be a rather crude
approach  for a weakly bound nucleus such as $^{11}$Be.
However,  we feel that it is, nevertheless,
good enough to reveal the importance of the PP

We found that only PQPRM reproduces satisfactorily the experimental ordering
of the lowest four levels in $^{13}$C with the correct value of
deformation $\beta=-0.6$.
Neither the PRM nor the QPRM could accomplish this.
 The  pairing
strongly reduces the coupling between the $0^+$ and the $2^+$
 core states for the negative parity states $1/2^-$ and $3/2^-$.
As a consequence,  within the  QPRM and  PQPRM
the ground state of $^{13}$C turns out
to be basically
the single-particle $1p_{1/2}$ state.
Our positive parity states $1/2^+$ and $5/2^+$, on the other hand,
are more collective when compared to results of the previous works.
Both effects seems to go in right  direction
and make the PQRPM to account for the experimental spectroscopic
factors of low lying positive and negative parity states, and for the
dipole magnetic moment of the ground state.

Similar calculations were done for one-neutron halo nucleus
$^{11}$Be. It is found that the more likely structure of the core,
$^{10}$Be, required to reproduce
the lowest three states in $^{11}$Be, is that of a vibrator. Moreover,
we feel that the appropriate model for
the experimental value of the vibrational length, $\beta=0.6$, is
again the PQPVM.
 As before, we found that inclusion of pairing makes the
negative-parity states less collective and positive parity states more
collective compared to earlier works. Also here the spectroscopic
factors and the dipole magnetic moment are well reproduced  by the PQPVM.

In summary, the inclusion of the pairing interaction and of the concomitant
Pauli Principle is imperative  not only in the case of heavy odd-mass
carbon isotopes but also in the core-coupling models of $^{13}$C and
$^{11}$Be.  The important role played by
the particle number conservation in these relatively light and/or exotic
nuclei has been confirmed as well.

\acknowledgments

Authors would like to acknowledge the partial
support of Brazilian agencies
Conselho Nacional de Ci\^encia e
Tecnologia (CNPq) and Funda\c{c}\~ao de Amparo \`a Pesquisa do
Estado de S\~ao Paulo (FAPESP).
A.R.S. acknowledges support received from Funda\c{c}\~ao de Amparo \`a Pesquisa do
Estado do Rio de Janeiro (FAPERJ).

\end{document}